\documentclass[twocolumn,pra,aps,showpacs,superscriptaddress,amsmath,amssymb]{revtex4}

\usepackage{graphicx}% Include figure files
\usepackage{dcolumn}% Align table columns on decimal point
\usepackage{bm}% bold math

\newcommand{\ket}[1]{| #1 \rangle}

\newcommand{\Hint}{H_{\mathrm{int}}}
\newcommand{\bara}{{\bar{a}}}
\newcommand{\barb}{{\bar{b}}}
\newcommand{\barc}{{\bar{c}}}
\newcommand{\bard}{{\bar{d}}}

\newcommand{\ee}{{\mathrm{e}}}
\newcommand{\ri}{{\mathrm{i}}}
\newcommand{\dd}{{\mathrm{d}}}

\newcommand{\dps}{\displaystyle }
\newcommand{\OO}{{\mathrm{O}}}

\begin{document}

\preprint{}

\title{Adiabatic approximation, Gell-Mann and Low theorem and
  degeneracies: \\ A pedagogical example}

\author{Christian Brouder}
\affiliation{
  Institut de Min\'eralogie et de Physique des Milieux Condens\'es,
  CNRS UMR 7590, Universit\'es Paris 6 et 7, IPGP, 140 rue de Lourmel,
  75015 Paris, France.
}

\author{Gabriel Stoltz}
\altaffiliation[Also at ]{
  Institut de Min\'eralogie et de Physique des Milieux Condens\'es,
  CNRS UMR 7590, Universit\'es Paris 6 et 7, IPGP, 140 rue de Lourmel,
  75015 Paris, France.
}
\affiliation{
  Universit\'e Paris Est, CERMICS, Projet MICMAC ENPC -
  INRIA, 6 \& 8 Av. Pascal, 77455 Marne-la-Vall\'ee Cedex 2, France
}
\author{Gianluca Panati}
\affiliation{
Department of Mathematics,
University La Sapienza,
Piazzale Aldo Moro, 2,
00185 Roma, Italie
}

\date{\today}

\begin{abstract}
We study a simple system described by a $2 \times 2$ Hamiltonian and the
evolution of the quantum states under the influence of a
perturbation. More precisely, when the initial Hamiltonian is not degenerate, 
we check analytically the validity of
the adiabatic approximation and verify that, even if the evolution operator 
has no limit for adiabatic switchings, the Gell-Mann and
Low formula allows to follow the evolution of eigenstates.  
In the degenerate case, for generic initial eigenstates, 
the adiabatic approximation (obtained by two
different limiting procedures) is either useless
or wrong, and the Gell-Mann and Low formula does not hold. We show
how to select initial states in order to avoid such failures.
\end{abstract}

%31.15am     Many-body perturbation calculation
%11.10-z     Field theory

\maketitle

%--------------------------------------------
%
%            INTRODUCTION
%
%--------------------------------------------

\section{Introduction}

Adiabatic switching is a crucial ingredient of many-body theory.
It provides a way to express
the eigenstates of a Hamiltonian $H=H_0+H_1$ in terms
of the eigenstates of $H_0$. 
Its basic idea is to switch very slowly the interaction
$H_1$, \textit{i.e.} to transform $H$ into a time-dependent 
Hamiltonian $H(t) = H_0 + \ee^{-\varepsilon|t|} H_1$
where the small parameter $\varepsilon > 0$ eventually vanishes.
Under the influence of $H(t)$, an eigenstate 
$\ket{\Upsilon_0}$ of $H_0$
becomes a time-dependent wavefunction 
$\ket{\Psi_\varepsilon(t)}$ and it might be expected that
an eigenstate of $H = H(0)$ is obtained by taking
the limit of $\ket{\Psi_\varepsilon(0)}$ when
$\varepsilon$ tends to zero.
It turns out that this naive expectation is not justified because
$\ket{\Psi_\varepsilon(0)}$ has no limit when $\varepsilon\to0$. 
When the initial state $\ket{\Upsilon_0}$ belongs to a non 
degenerate eigenspace (isolated from the other eigensubspaces),
Gell-Mann and Low \cite{GellMann} solved the problem
by dividing out the divergence by a suitable factor. 
The ratio is called the Gell-Mann and Low wavefunction
and its convergence can be proved by using
the adiabatic theorem \cite{Nenciu}.

In the first part of this work, we present an exactly 
solvable $2 \times 2$ model that illustrates the fact that
the limit $\varepsilon\to0$ of $\ket{\Psi_\varepsilon(0)}$ does not exist.
The validity of the Gell-Mann and Low wavefunction
is shown by analytically calculating the corresponding
adiabatic approximation.

It was realized fifty years ago \cite{BH58} that 
the Gell-Mann and Low formula must sometimes be extended to the case of 
a degenerate initial state of $H_0$.
This happens in many practical situations, for instance when the
system contains unfilled shells.
This problem has been discussed 
in several fields, including nuclear physics \cite{KuoOsnes},
solid state physics \cite{Esterling},
quantum chemistry \cite{MukherjeeGML}
and atomic physics \cite{Lindgren3}.
In most cases, it is assumed that there
is \emph{some} eigenstate $\ket{\Upsilon_0}$
for which the Gell-Mann and Low formula holds.
More precisely, if $\mathcal{V}_0$ is the vector space
generated by the eigenstates of $H_0$
associated with the degenerate ground state energy $E_0$, then
the claim is that there exists some initial state $\ket{\Upsilon}$ in
$\mathcal{V}_0$ whose time evolved state
$\ket{\Psi_\varepsilon(0)}$ is (up to a divergent phase)
an eigenstate of $H$.
To see that it is not possible to choose any element
of $\mathcal{V}_0$ as initial state, let us forget for a moment the
divergent phase and consider a perturbation
$H_1$ that splits the degeneracy and two initial 
states $\ket{\Upsilon^1}$ and $\ket{\Upsilon^2}$ whose
time evolutions give rise to two eigenstates of $H$,
denoted by $\ket{\Psi^1_\varepsilon(0)}$ and $\ket{\Psi^2_\varepsilon(0)}$,
with different energies $E_1$ and $E_2$. By linearity,
the initial state $\ket{\Upsilon^1}+\ket{\Upsilon^2}$ 
evolves into 
$\ket{\Psi^1_\varepsilon(0)}+\ket{\Psi^2_\varepsilon(0)}$
which is not an eigenstate of $H$ because
$E_1\not=E_2$. Therefore, 
$\ket{\Upsilon^1}+\ket{\Upsilon^2}$ is not a proper initial state.

When the initial state of $H_0$ is degenerate, we show that, even
for the very simple $2 \times 2$ model considered here, 
the Gell-Mann and Low wavefunction does not have a limit for
almost all initial conditions.
We find however the specific initial states that lead 
to convergent wavefunctions. 
For the application of many-body methods to degenerate
systems, it is crucial to find a way to select
the proper initial states.
%Based on the careful analysis of the simple $2 \times 2$ model used
%here, we suggest a general method to choose the initial state so that
%the Gell-Mann and Low wavefunction has a limit when $\varepsilon \to 0$.

%--------------------------------------------
%
%           CAS SANS DEGENERESCENCE
%
%--------------------------------------------

\section{Non-degenerate initial states}

Two-dimensional matrices are the simplest non-trivial models that can
be considered in quantum physics, and were indeed already used as toy-models
for many-body theory studies (see Ref.~\onlinecite{Ellis}
and references therein). However, it is the first
time to our knowledge that the evolution operator and its adiabatic
approximation are calculated explicitly for such a model. 

We consider the two-dimensional system described by the Hamiltonian
$H=H_0+H_1$ with
\begin{equation}
\label{eq:2D_model}
H_0 = \left( \begin{array}{cc} \mu - \delta & 0 \\
   0 & \mu+\delta \end{array}\right),
\quad
H_1 = \left( \begin{array}{cc} 0 & x \\
   x & 0 \end{array}\right).
\end{equation}
We assume that $\delta\not=0$, so that the initial state is non
degenerate. Without restriction, it can further be assumed that
$\delta > 0$.

According to the usual treatment of adiabatic switching
\cite{Fetter,Gross}, it is convenient to transform the
Schr\"odinger equation for the time-dependent Hamiltonian 
$H(t)=H_0+\ee^{-\varepsilon |t|} H_1$ to the interaction picture.
The dependence of the operators and wavefunctions on the
parameter $\varepsilon>0$ will be implicit in the sequel.
If we denote by $\Psi_S(t)$ a solution of the Schr\"odinger equation
\[
\ri \frac{\partial \Psi_S(t)}{\partial t} = H(t) \Psi_S(t),
\]
the wavefunction in the interaction picture is defined as
\[
\Psi(t)=\ee^{\ri H_0 t}\Psi_S(t).
\]
It satisfies 
\[
\ri\frac{\partial\Psi(t)}{\partial t} = \Hint(t)\Psi(t), 
\]
subject to the boundary condition 
$\Psi(-\infty) = \Upsilon$, where $\Upsilon$ may be an eigenstate of $H_0$.
The Hamiltonian $\Hint(t)$ is the Hamiltonian $H(t)$ in the
interaction picture
\[
\Hint(t) = \ee^{-\varepsilon|t|} \, \ee^{\ri H_0t} \, H_1 \, 
\ee^{-\ri H_0t}.
\]
In the simple case~\eqref{eq:2D_model} considered here,
this operator reads
\begin{equation}
  \Hint(t) = x \, \ee^{-\varepsilon|t|} 
  \left( \begin{array}{cc} 
    0 & \ee^{-2\ri\delta t} \\
    \ee^{2\ri\delta t} & 0 
  \end{array}\right).
\end{equation}

Instead of using wavefunctions, it is customary
to work with the evolution matrix $U(t)$ such that
$\Psi(t)=U(t)\Upsilon$.
We refer to \cite{ReedSimonII} for sufficient conditions on the time
dependent Hamiltonian $H_\mathrm{int}(t)$ in order to ensure the
existence of the unitary propagator $U(t)$.

%----------------------------------
%       Operateur d'evolution 
%----------------------------------
\subsection{The evolution operator: analytic solution and limiting behavior}
\label{sec:evolution}

We give in this section the analytic expression of the evolution
matrix. This operator is the solution of the Schr\"odinger equation
in the interaction picture
\begin{equation}
  \label{eq:Schr_evolution}
  \ri \frac{\dd U(t)}{\dd t} = \Hint(t) U(t),
\end{equation}
with the boundary condition $U(-\infty) = \mathrm{Id}$.
In the sequel, we consider $t \le 0$, so that
the switching function is $\ee^{\varepsilon t}$.
We also assume that $x > 0$.
Denoting the matrix elements of $U(t)$ by
\begin{eqnarray*}
U(t) = \left( \begin{array}{cc} a(t) & b(t) \\
   c(t) & d(t) \end{array}\right),
\end{eqnarray*}
eq.~\eqref{eq:Schr_evolution} is
equivalent to the system of equations
\begin{equation}
  \label{cprime}
  \left \{ \begin{array}{ccl}
  \ri a' & = & x \, \ee^{-\ri(2\delta+\ri\varepsilon) t} \, c, \\
  \ri c' & = & x \, \ee^{ \ri(2\delta-\ri\varepsilon) t} \, a,  \\
  \ri b' & = & x \, \ee^{-\ri(2\delta+\ri\varepsilon) t} \, d, \\
  \ri d' & = & x \, \ee^{ \ri(2\delta-\ri\varepsilon) t} \, b,
  \end{array} \right.
\end{equation}
with the boundary conditions
\[
a(-\infty) = d(-\infty) = 1, \quad b(-\infty) = c(-\infty) = 0.
\]
Since the functions $c,d$ are independent of the functions $a,b$, and
satisfy equations of the same form as for $a,b$, it is enough to solve
the first two equations of the above system.

%------------ solution pour a -----------
\subsubsection{Solution of the equation for the unknown function $a$}

By eliminating the function $c$ in eq.~(\ref{cprime}), the second order
evolution equation
\begin{eqnarray}
a'' + (2\ri\delta - \varepsilon) a' + x^2 \ee^{2\varepsilon t} a = 0
\label{eqa}
\end{eqnarray}
is obtained.
This equation can be solved by using standard 
techniques~\cite[Eq.~(23), p.~442]{Kamke}: If we rewrite the
unknown function $a$ as
$a(t) = \ee^{-\ri(\delta+\ri\varepsilon/2) t}Z_\nu(x \ee^{\varepsilon t}/\varepsilon)$,  
and introduce the variable $s = x\ee^{\varepsilon t}/\varepsilon$, 
eq.~(\ref{eqa}) becomes the Bessel equation for $Z_\nu(s)$, with
\begin{equation}
  \label{eq:nu}
  \nu = 1/2-\ri\delta/\varepsilon.
\end{equation}
Since $J_\nu$ and $J_{-\nu}$
are two independent solutions of the Bessel equation
when $\nu$ is not an integer \cite{Abramowitz},
the function $a(t) = \bara(s)$ has the general form
\begin{equation}
  \label{eq:general_form}
  \bara(s) = \left(\frac{\varepsilon s}{x}\right)^\nu 
  \left ( C_1 J_\nu(s) + C_2  J_{-\nu}(s) \right),
\end{equation}
where the constants $C_1,C_2$ are determined by the boundary
conditions at $s = 0$ (\textit{i.e} in the limit $t \to -\infty$).
Since $a(-\infty) = \bara(0) = 1$, 
and by the series expansion \cite[Eq.~(9.1.10)]{Abramowitz} %p.~360
\[
J_\nu(s) = \left(\frac{s}{2}\right)^\nu \sum_{k=0}^\infty
   \frac{(-s^2/4)^k}{k! \, \Gamma(\nu+k+1)},
\]
it follows $C_1 = 0$ and $C_2 = 2^{-\nu} \Gamma(1-\nu)$.

%------------ solution pour autres -----------
\subsubsection{Solution of the equation for the other unknown functions}

The function $c(t) = \barc(s)$ can be obtained from the expression of
$a(t) = \bara(s)$, relying on the first equation in the system
\eqref{cprime}, rewritten in the $s$ variable as 
$\barc(s) = \ri(\varepsilon s/x)^{1-2\nu}\bara'(s)$.
Since $(s^\nu J_{-\nu}(s))'=-s^\nu J_{1-\nu}(s)$ 
\cite[Eq.~(9.1.30)]{Abramowitz}, %p.~361
it holds
\[
\barc(s) = - \ri C_2 \left(\frac{\varepsilon}{x}\right)^{1-2\nu} 
s^{1-\nu} J_{1-\nu}(s).
\]
The functions $b$ and $d$ satisfy the same equations
as $a$ and $c$, respectively, but with different boundary
conditions.
With the above notation, it can easily be checked that
$b$ is also a combination of Bessel 
functions as given by \eqref{eq:general_form}, but with 
$C_1=-\ri(\varepsilon/x)^{2\nu-1}2^{\nu-1}\Gamma(\nu)$
and $C_2=0$.

%------------ proprietes -----------
\subsubsection{Properties of the solution}

In view of the above results, the problem \eqref{cprime} has the
analytic solution
\begin{eqnarray*}
a(t) &=& C_2 \, \left(\frac{x}{\varepsilon}\right)^\nu \ee^{\varepsilon\nu t}
   J_{-\nu}\left(\frac{x \ee^{\varepsilon t}}{\varepsilon}\right),\\
b(t) &=& C_1 \, \left(\frac{x}{\varepsilon}\right)^\nu \ee^{\varepsilon\nu t}
   J_\nu\left(\frac{x \ee^{\varepsilon t}}{\varepsilon}\right),\\
c(t) &=& -\ri C_2 \, \left(\frac{x}{\varepsilon}\right)^\nu
     \ee^{\varepsilon(1-\nu) t}
     J_{1-\nu}\left(\frac{x \ee^{\varepsilon t}}{\varepsilon}\right),\\
d(t) &=& \ri C_1 \, \left(\frac{x}{\varepsilon}\right)^\nu
     \ee^{\varepsilon(1-\nu) t}
   J_{\nu-1}\left(\frac{x \ee^{\varepsilon t}}{\varepsilon}\right),
\end{eqnarray*}
with 
\[
C_1 = -\ri(\varepsilon/x)^{2\nu-1}2^{\nu-1}\Gamma(\nu), 
\quad 
C_2 = 2^{-\nu} \Gamma(1-\nu),
\]
and where $\nu$ is defined in \eqref{eq:nu}.
As a consistency check, it is possible to verify that 
the matrix $U$ is unitary. This follows from
$|C_1|^2=|C_2|^2=\Gamma(\nu)\Gamma(1-\nu)/2=\pi/(2\sin\nu\pi)$
and
$J_{\nu}(s)J_{1-\nu}(s)+J_{-\nu}(s)J_{\nu-1}(s)
=2\sin(\nu\pi)/(\pi s)$ (see \cite[Eq.~(9.1.15)]{Abramowitz}).

The series expansion of the Bessel functions gives
\begin{eqnarray*}
\bara(s) & = &
1 + \sum_{k=1}^\infty \frac{(-s^2/4)^k }{k! \, \prod_{j=1}^{k} (j-\nu)},\\
\barb(s) & = & \ri\ee^{-2\ri\delta t} \frac{s}{2}
\sum_{k=0}^\infty \frac{(-s^2/4)^k }{k! \, \prod_{j=1}^{k+1} (j+\nu-1)},\\
\barc(s) & = & -\ri\ee^{2\ri\delta t} \frac{s}{2}
\sum_{k=0}^\infty \frac{(-s^2/4)^k }{k! \, \prod_{j=1}^{k+1} (j-\nu)},\\
\bard(s) & = &
1 + \sum_{k=1}^\infty \frac{(-s^2/4)^k }{k! \, \prod_{j=1}^{k} (j+\nu-1)}.
\end{eqnarray*}
Note that these solutions are valid for all values of $x$, $\delta\not=0$
and $\varepsilon>0$.
We have the symmetries $a(t,-\delta)=d(t,\delta)$ and
$c(t,-\delta)=-b(t,\delta)$, $a(t)^*=d(t)$ and $b(t)^*=c(t)$.
Moreover, $a(t)$ and $d(t)$ are even in $x$, 
while $b(t)$ and $c(t)$ are odd in $x$.

%------------ limites -----------
\subsubsection{Limiting behavior of the evolution operator}

\begin{figure}[ht]
\begin{center}
\includegraphics[width=8cm]{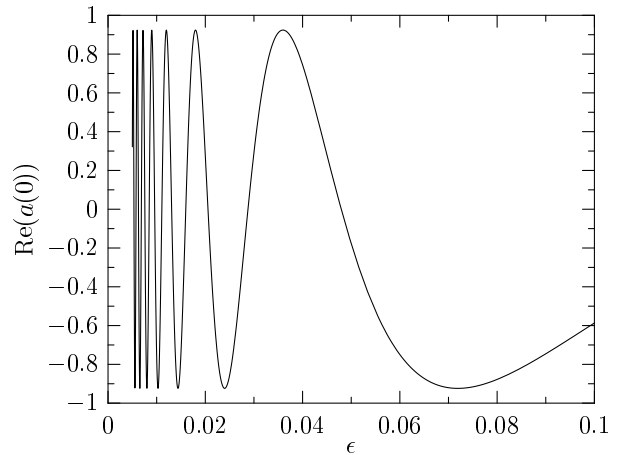}
\caption{\label{fig:are}
Real part of $a(0)$ as a function of $\varepsilon$
for $x = \delta = 1$ and $\mu = 0$. For clarity,
the region $x<0.005$ is not plotted.
}
\end{center}
\end{figure}

Figure~\ref{fig:are} illustrates the divergence of $a(0)$ as
a function of $\varepsilon$. The above analytic expressions therefore
show that the evolution operator does not have a limit when
$\varepsilon \to 0$. Rather, as illustrated by Fig.~\ref{fig:are}, a
strongly oscillatory behavior is observed. This is the reason why the
phase factor has to be cancelled out by considering a renormalized
wavefunction in the Gell-Mann and Low fashion, as explained in the
next section.

%---------------------------
%   Approx adiabatique  
%---------------------------
\subsection{The Gell-Mann and Low wavefunction}

Rigorous proofs of the Gell-Mann and Low formula rely 
on the adiabatic approximation. Therefore, we first compute the
adiatic approximation for the problem \eqref{eq:2D_model}, and then
the Gell-Mann and Low wavefunctions.

%------------ approx adiabatique -----------
\subsubsection{The adiabatic approximation}
\label{adiabsect}

The adiabatic approximation consists in approximating the evolution
operator in the interaction picture $U(t)$ by
the adiabatic operator in the interaction picture:
\begin{equation}
  \label{eq:adiabatic}
  U_a(t) = \ee^{\ri t H_0} A(\tau) \overline{\Phi}_\varepsilon(\tau),
\end{equation}
with
\[
\overline{\Phi}_\varepsilon(\tau) =  \lim_{\tau_0\to -\infty}
\Phi_\varepsilon(\tau,\tau_0)\ee^{-\ri t_0H_0},
\]
where $\tau=\varepsilon t < 0$ and $\tau_0=\varepsilon t_0$ \cite{Kato50,Messiah}.
The unitary matrices $A$ and $\Phi_\varepsilon$
are calculated from the eigenprojectors of $H(\tau)=H_0+\ee^{\tau}
H_1$, as explained below 
(see formulas \eqref{Adetau} and \eqref{eq:matrix_Phi}).

The eigenvalues of $H(\tau)$ are
\[
e_\pm(\tau)=\mu\pm\lambda, 
\qquad
\lambda=\sqrt{\delta^2+x^2\ee^{2\tau}}.
\]
%The associated eigenvectors are respectively
%\begin{eqnarray*}
%v_+(\tau) &=& \left(\begin{array}{c}
%  \sqrt{\frac{\lambda-\delta}{2\lambda}} \\
%  \frac{x\ee^\tau}{\sqrt{2\lambda(\lambda-\delta)}}
%\end{array}\right),
%\quad
%v_-(\tau) = \left(\begin{array}{c}
%  \sqrt{\frac{\lambda+\delta}{2\lambda}} \\
%  -\frac{x\ee^\tau}{\sqrt{2\lambda(\lambda+\delta)}}
%\end{array}\right).
%\end{eqnarray*}
With the notation, $y = x\ee^\tau/\delta$,
the corresponding eigenprojectors read respectively
\[
P_\pm(\tau) = \frac{1}{2\sqrt{1+y^2}} \left(\begin{array}{cc}
  \sqrt{1+y^2}\mp 1 & \pm y \\
  \pm y & \sqrt{1+y^2}\pm 1
\end{array}\right).
\]
Defining the Hermitian matrix \cite{Messiah}
\[
K(\tau) = \ri\sum_{\sigma=\pm} \frac{\dd P_\sigma(\tau)}{\dd\tau}
P_\sigma(\tau)
=  \frac{\ri y}{2(1+y^2)} \left(\begin{array}{cc}
  0 & 1 \\
  -1 & 0
\end{array}\right),
\]     
the matrix $A$ appearing in \eqref{eq:adiabatic} satisfies
the equation 
\begin{equation}
  \label{eq:A}
  A'(\tau) = -\ri K(\tau)A(\tau),
\end{equation}
with the boundary condition $A(-\infty) = \mathrm{Id}$. 
The unique solution is
\begin{equation}
  \label{Adetau}
  A(\tau) = \frac{1}{\sqrt{2}}\left(\begin{array}{cc}
    \sqrt{1+\alpha} & \sqrt{1-\alpha} \\
    -\sqrt{1-\alpha} & \sqrt{1+\alpha}
  \end{array}\right),
\end{equation}
with $\alpha = (1+y^2)^{-1/2} \in [0,1]$.

The phase matrix $\Phi_\varepsilon(\tau,\tau_0)$ is
obtained from the integral of the eigenvalues:
\[
\Phi_\varepsilon(\tau,\tau_0) =  
  \sum_{\sigma = \pm} \exp\left(-\frac{\ri}{\varepsilon}\int_{\tau_0}^\tau
  e_\sigma(\tau')\dd \tau' \right) P_\sigma(\tau_0).
\]
For large negative $\tau_0$, up to
$\OO(\ee^{\tau_0})$ terms coming from the approximation
$P_\sigma(\tau_0) \simeq P_\sigma(-\infty)$,
\begin{equation}
  \label{eq:matrix_Phi}
  \Phi_\varepsilon(\tau,\tau_0) = \ee^{-\ri\mu(t-t_0)}
  \left(\begin{array}{cc}
    \ee^{\ri\phi(\tau,\tau_0)/\varepsilon} & 0 \\
    0 & \ee^{-\ri\phi(\tau,\tau_0)/\varepsilon}
  \end{array}\right),
\end{equation}
where
\begin{eqnarray*}
\phi(\tau,\tau_0) &=& \int_{\tau_0}^\tau 
   \sqrt{\delta^2+x^2\ee^{2\sigma}} \dd\sigma
= F(\tau)-F(\tau_0),
\end{eqnarray*}
with
\begin{eqnarray*}
F(\tau) &=& \sqrt{\delta^2+x^2\ee^{2 \tau}}
  -\frac{\delta}{2}
  \log\left(\frac{\sqrt{\delta^2+x^2\ee^{2\tau}}+\delta}
      {\sqrt{\delta^2+x^2\ee^{2\tau}}-\delta}\right).
\end{eqnarray*}
Note that $F(\tau_0)$ has no finite limit for $\tau_0 \to -\infty$.
However, because the computations are done in the 
interaction picture, the phase matrix 
$\Phi_\varepsilon(\tau,\tau_0)$ is multiplied
by the unitary operator $\ee^{-\ri t_0 H_0}$, 
and the so-obtained operator has a limit when $\tau_0 \to -\infty$. Indeed,
\begin{eqnarray*}
  \overline{\Phi}_\varepsilon(\tau) 
  &=& \lim_{t_0\to -\infty}
  \Phi_\varepsilon(\tau,\tau_0)\ee^{-\ri H_0 t_0} \\
  &=& \ee^{-\ri\mu \tau/\varepsilon} 
  \left(\begin{array}{cc}
    \exp(\ri\phi(\tau)/\varepsilon) & 0 \\
    0 & \exp(-\ri\phi(\tau)/\varepsilon)
  \end{array}\right),
\end{eqnarray*}
where
\begin{eqnarray*}
  \phi(\tau) 
  &=& \lim_{t_0\to -\infty} 
  \mu\tau_0 + \phi(\tau,\tau_0)-(\mu-\delta) \tau_0 \\
  &=& F(\tau) - \delta\left(1-\log\frac{2\delta}{x}\right).
\end{eqnarray*}

At $t=0$, we obtain the adiabatic approximation
\begin{eqnarray*}
U_a(0) &=& \frac{1}{\sqrt{2}}
  \left(\begin{array}{cc}
    \ee^{\ri\phi(0)/\varepsilon} \sqrt{1+\alpha} & 
    \ee^{-\ri\phi(0)/\varepsilon} \sqrt{1-\alpha} \\
    -\ee^{\ri\phi(0)/\varepsilon} \sqrt{1-\alpha} & 
    \ee^{-\ri\phi(0)/\varepsilon} \sqrt{1+\alpha}
  \end{array}\right),
\end{eqnarray*}
where $\alpha$ is now $(1+x^2/\delta^2)^{-1/2}$.

%------------ quality check -----------
\subsubsection{Quality of the adiabatic approximation}

The rigorous adiabatic theorem
(see Ref.~\onlinecite{Teufel} for a recent account)
states that the adiabatic evolution operator $U_a$ defined by 
Eq.~\eqref{eq:adiabatic} is such that 
\[
\sup_{t \leq 0} \| U(t)-U_a(t) \| \leq C \varepsilon,
\]
for some constant $C$.
In particular, for $t=0$,
$\| U(0)-U_a(0) \| \leq C \varepsilon$.
Figure 2 shows that this result is indeed verified in the case
considered here, and allows to give a numerical estimate 
of the constant $C$.

\begin{figure}[ht]
\begin{center}
\includegraphics[width=8cm]{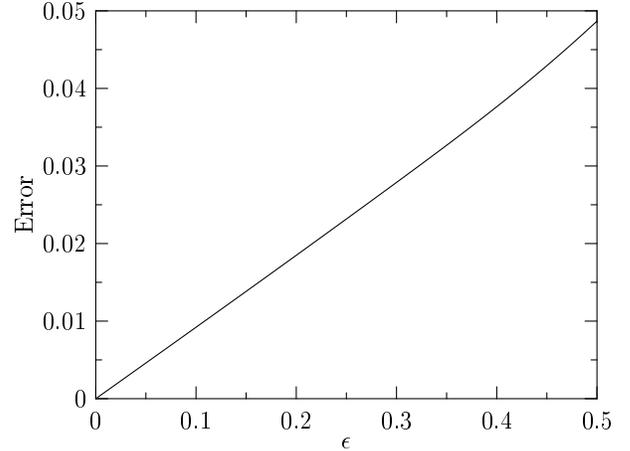}
\caption{\label{fig:des}
Modulus of the difference between $a(0)$ and its
adiabatic approximation as a function of $\varepsilon$
for $x = \delta = 1$ and $\mu = 0$.
}
\end{center}
\end{figure}

%------------ Gell-Mann and Low -----------
\subsubsection{The Gell-Mann and Low limit}
\label{sec:GLL}

The Gell-Mann and Low formula describes how an eigenvector of the
reference (unperturbed) Hamiltonian $H_0$ evolves under an added
perturbation $H_1$. Starting from one of the initial 
eigenstates of $H_0$, namely
\begin{equation}
  \label{eq:def_initial_states}
  \ket{\Upsilon^1} = \begin{pmatrix}
    1 \\ 0
  \end{pmatrix}, 
  \qquad 
  \ket{\Upsilon^2} = \begin{pmatrix}
    0 \\ 1
  \end{pmatrix},
\end{equation}
the associated Gell-Mann and Low wavefunctions 
\[
\ket{\Psi^i(0)} = \lim_{\varepsilon \to 0} 
\frac{U(0) \ket{\Upsilon^i}}{\langle\Upsilon^i| U(0) \ket{\Upsilon^i} }
= \lim_{\varepsilon \to 0} 
\frac{U_a(0) \ket{\Upsilon^i}}{\langle\Upsilon^i| U_a(0) \ket{\Upsilon^i} }
\] 
can be computed using \eqref{eq:adiabatic}.
They read respectively 
\[
\left( \begin{array}{c} 
  1 \\[5pt] \dps \frac{a_{21}(0)}{a_{11}(0)}
\end{array}\right) =
\left( \begin{array}{c} 
  1 \\[5pt] \dps -\frac{x}{\sqrt{x^2+\delta^2}+\delta}
\end{array}\right), 
\] 
and 
\[
\left( \begin{array}{c} 
  \dps \frac{a_{12}(0)}{a_{22}(0)} \\[10pt] 1
\end{array}\right) =
\left( \begin{array}{c} 
  \dps \frac{\sqrt{x^2+\delta^2}+\delta}{x} \\[10pt] 1
\end{array}\right),
\]
where $a_{ij}(\tau)$ are the matrix elements
of $A(\tau)$ given by (\ref{Adetau}). 
It is easy to check that these vectors are indeed 
eigenstates of $H$, for the eigenvalues $\mu-\sqrt{x^2+\delta^2}$ and
$\mu+\sqrt{x^2+\delta^2}$ respectively.
These eigenvalues can be obtained by the
energy-shift formula~\cite[p.~200]{Gross}.
The difference $\Delta E$ between the energy of an eigenstate 
of $H_0+H_1$ and of the corresponding eigenstate
$|\Upsilon_0\rangle$ of $H_0$ is
\begin{eqnarray*}
\Delta E &=& \lim_{\varepsilon\to0} \ri
\varepsilon x\frac{\dd}{\dd x}\log\langle\Upsilon_0|U(0)|\Upsilon_0\rangle
\\&=&
\lim_{\varepsilon\to0} \ri \varepsilon x\frac{\dd}{\dd
x}\log\langle\Upsilon_0|U_a(0)|\Upsilon_0\rangle.
\end{eqnarray*}
We find indeed that $\Delta E= \delta-\sqrt{x^2+\delta^2}$
for $|\Upsilon^1\rangle$ and
$\Delta E= -\delta+\sqrt{x^2+\delta^2}$
for $|\Upsilon^2\rangle$.

%------------------------------------------
%
%             DEGENERATE CASE
%  
%------------------------------------------

\section{Degenerate case}

Initial degenerate states for the model \eqref{eq:2D_model} are obtained when 
$\delta = 0$. In this case, the evolution operator can still be
computed analytically, and actually has a simpler expression than the
one obtained in Section~\ref{sec:evolution} in the non-degnerate case: 
\begin{equation}
  \label{Utdeg}
  U(t) = \left( \begin{array}{cc} 
    \cos (x \ee^{\varepsilon t}/\varepsilon) &
    -\ri \sin (x \ee^{\varepsilon t}/\varepsilon) \\
    -\ri \sin (x \ee^{\varepsilon t}/\varepsilon) & 
    \cos (x \ee^{\varepsilon t}/\varepsilon)
  \end{array}\right).
\end{equation}
In particular,
\[
U(0) = \left( \begin{array}{cc} 
  \cos (x/\varepsilon) & -\ri \sin (x/\varepsilon) \\
  -\ri \sin (x/\varepsilon) & \cos (x/\varepsilon).
\end{array}\right).
\]

%-----------------------------
%       FAILURES
%-----------------------------

\subsection{Failure of the adiabatic approximation and the Gell-Mann
  and Low formula}
\label{sec:failure}

%------------- adiabatic ---------
\subsubsection{Adiabatic approximation}

There are two ways to calculate the adiabatic approximation,
depending on the order of the limits $\delta\to0$
and $\tau_0\to -\infty$.

If we first carry out the limit $\delta\to 0$, the computation of Section
\ref{adiabsect} can be repeated by starting from the Hamiltonian
$H_0$ with $\delta=0$. The eigenvalues corresponding to $H(\tau)$
are $\mu\pm x \ee^\tau$, and the eigenprojectors
are constant:
\[
P_\pm(\tau) = \frac{1}{2} \left(\begin{array}{cc}
  1 & \pm 1\\
  \pm 1 & 1
\end{array}\right).
\]
Thus, $K(\tau) = 0$, so that $A(\tau)$ is constant
and equal to its boundary value $A(\tau) = \mathrm{Id}$.
The computation of $\overline{\Phi}_\varepsilon(\tau)$
is therefore straighforward and leads to the adiabatic approximation
\[
U_a(t) = U(t), 
\]
where $U(t)$ is given by Eq.~(\ref{Utdeg}).
The adiabatic approximation is therefore exact but,
as we shall see, the Gell-Mann and Low wavefunction has
no limit for the initial states $\Upsilon^1$ and $\Upsilon^2$
when $\varepsilon \to 0$.

Taking first the limit $\tau_0\to -\infty$ amounts
to take the limit $\delta \to 0$
in the definition of the adiabatic evolution operator
\eqref{eq:adiabatic}. For the operator $A$,
the limit is
\begin{equation}
  \label{Ad=0}
  A(\tau) = \frac{1}{\sqrt{2}}
  \left(\begin{array}{cc}
    1 &  1 \\
    -1 &  1
  \end{array}\right),
\end{equation}
for any $ \tau \leq 0$. Notice that the boundary condition at
$\tau=-\infty$ for the equation \eqref{eq:A} on $A$ 
is therefore not satisfied.
For the matrix $\Phi_\varepsilon(\tau)$, the limit is
\[
\Phi_\varepsilon(\tau) =
\ee^{-\ri\mu t} \left(\begin{array}{cc}
  \exp(\ri x \ee^\tau/\varepsilon) & 0 \\
  0 & \exp(-\ri x \ee^\tau/\varepsilon)
\end{array}\right).
\]
The limit of the adiabatic evolution operator reads now
\begin{eqnarray}
U_a(t) = \frac{1}{\sqrt{2}}
\left(\begin{array}{cc}
  \exp(\ri x\ee^{\varepsilon t}/\varepsilon) &
  \exp(-\ri x\ee^{\varepsilon t}/\varepsilon) \\
  - \exp(\ri x\ee^{\varepsilon t}/\varepsilon) & 
  \exp(-\ri x\ee^{\varepsilon t}/\varepsilon) 
\end{array}\right).
\label{adiabatic-limit-2}
\end{eqnarray}
It is however not an approximation of $U(t)$.

The discrepancy between the two approaches clearly shows that
the two limits $\delta \to 0$ and $\tau_0\to -\infty$ do not
commute for degenerate systems.

%--------------- GM & L ----------------
\subsubsection{Non-validity of the general Gell-Mann and Low formula}
\label{sec:nonval_GLL}

Consider a given initial state $\ket{\Upsilon_0}$, which is an
eigenvector associated with the eigenvalue $\mu$. Since the
corresponding eigenspace is two-dimensional, there seems to be
some arbitrariness in the choice of the initial state.
However, we shall see that most initial states lead
to divergent Gell-Mann and Low wavefunctions.
If for instance $\ket{\Upsilon_0} = \ket{\Upsilon^1}
= \begin{pmatrix} 1 \\ 0 \end{pmatrix}$
as in Section~\ref{sec:GLL}, the Gell-Mann and Low wavefunction
\[
\frac{U(0)\ket{\Upsilon_0}}{\langle\Upsilon_0|U(0)|\Upsilon_0\rangle}
= \left( \begin{array}{c} 
  1 \\
  -\ri \tan (x/\varepsilon) 
\end{array}\right),
\]
has no limit when $\varepsilon \to 0$.
This shows that the Gell-Mann and Low formula fails
for the initial state $\ket{\Upsilon^1}$. It fails also for $\ket{\Upsilon^2}$.

Similarly, the energy shift would be given by
the limit for $\varepsilon\to0$ of
\begin{eqnarray*}
\Delta E &=& \ri
\varepsilon x\frac{\dd}{\dd x} \log\langle \Upsilon^1 | U(0) |\Upsilon^1\rangle
= - \ri x \tan(x/\varepsilon).
\end{eqnarray*}
But this has no limit when $x\not=0$.

The question is then whether there are some initial states for which
the Gell-Mann and Low wavefunction has a limit, and how those states can
be characterized.

%-------------------------------------
%     SELECTION OF RIGHT DIRECTIONS
%--------------------------------------

\subsection{Selection of the proper initial states}

In this section, we try to find initial states that are eigenstates
of $H_0$ and that lead to convergent Gell-Mann and Low 
wavefunctions.
%, as well as correct and convergent adiabatic
% approximations (for which all the $\varepsilon$ dependence is in some
% phase factor instead of a phase matrix). 

Any initial state $\ket{\Upsilon_0}$ for the model
\eqref{eq:2D_model} in the degenerate case
is of the general form (up to a trivial scaling changing this state into 
$-\ket{\Upsilon_0}$)
\[
\ket{\Upsilon_0} = \cos \theta \, \ket{\Upsilon^1} 
+ \sin \theta \, \ket{\Upsilon^2},
\]
where $0 \leq \theta < \pi$ and $\ket{\Upsilon^1},\ket{\Upsilon^2}$ are
defined in \eqref{eq:def_initial_states}.
Straightforward computations show that
\[
\frac{U(0)\ket{\Upsilon_0}}{\langle\Upsilon_0| U(0) |\Upsilon_0\rangle},
\]
has a limit if and only if $\theta = \pi/4$ or $\theta = 3\pi/4$.
This defines two proper initial states 
\[
\ket{\Upsilon_\pm} = \frac{1}{\sqrt{2}} 
\begin{pmatrix}
  1 \\ \pm 1
\end{pmatrix},
\]
with associated Gell-Mann and Low wavefunctions
\[
\frac{U(0)\ket{\Upsilon_\pm}}{\langle\Upsilon_\pm|U(0)|\Upsilon_\pm\rangle}
= \left( \begin{array}{c} 
  1 \\
  \pm 1 
\end{array}\right),
\]
which obviously have limits in the regime $\varepsilon = 0$.
Moreover, in the basis 
$(\ket{\Upsilon_-},\ket{\Upsilon_+})$,
the evolution operator becomes
\begin{eqnarray}
U(t) = \frac{1}{\sqrt{2}} \left(\begin{array}{cc}
  \exp(\ri x\ee^{\varepsilon t}/\varepsilon) & 
  \exp(-\ri x\ee^{\varepsilon t}/\varepsilon) \\
  - \exp(\ri x\ee^{\varepsilon t}/\varepsilon) & 
  \exp(-\ri x\ee^{\varepsilon t}/\varepsilon) 
\end{array}\right).
\end{eqnarray}
Similarly, the energy-shift formula gives now the correct result
$\Delta E_\pm = \pm x$.

Interestingly, the adiabatic operator given in
eq.~(\ref{adiabatic-limit-2}) is equal to the evolution
operator in the basis $(\ket{\Upsilon_-},\ket{\Upsilon_+})$.

%-------------------------------------
%
%           DISCUSSION
%
%-------------------------------------

\section{Conclusion}
The adiabatic theorem, which was recently questioned~\cite{Marzlin},
was investigated here by using an exactly solvable model.
Within this model, the adiabatic theorem is valid when the initial state
of the system is non-degenerate. When it is degenerate, our simple model
exhibits several problems that are generally present for an
arbitrary initial state: the Gell-Mann and
Low wavefunction does not converge, the
energy-shift formula is not valid and
the adiabatic approximation becomes ambiguous.
Within our model, all these problems are solved by 
properly choosing the initial state.

At least since Tolmachev \cite{TolmachevI}, it is conjectured that the Gell-Mann
and Low formula is valid for properly chosen initial states when the system is
degenerate. However, this conjecture has not been proved 
and no practical method 
was given to select the proper initial states. 
We intend to come back to this question in a forthcoming paper.

%-------------------------------------
%
%           BIBLIOGRAPHY
%
%-------------------------------------

%\bibliography{qed}

\end{document}